# Diversity legitimizes science: Holding basic research in the physical sciences accountable to the public

Kay T. Xia,[1]* Thayer L. Anderson,[2]† Phelan Yu[3]†

[1] Department of Chemistry, University of California, Berkeley, California 94720, United States. Current affiliation: Division of Chemistry and Chemical Engineering, Arthur Amos Noyes Laboratory of Chemical Physics, California Institute of Technology, Pasadena, California 91125, United States.

[2] Institute for Computational Medicine, Johns Hopkins University, Baltimore, Maryland 21218, United States.

[3] Division of Physics, Mathematics, and Astronomy, California Institute of Technology, Pasadena, California 91125, United States.

† These authors contributed equally

*K.T.X.: ktx@berkeley.edu

**Abstract**

The American scientific community is reeling from funding cuts and policy directives that will debilitate scientific research and education. The underlying hostilities fueling these attacks have intensified in recent years as the COVID-19 pandemic increased suspicion of scientific experts and the institutional embrace of diversity, equity, and inclusion (DEI) policies in 2020 prompted a backlash along longstanding political fault lines. Under the banner of anti-elitism, opponents of science and DEI have formed a coalition that sees attacks on higher education as a strategic means to achieve their political ends. While some of their arguments contain legitimate criticisms, academics must resist these attacks that seek to dismantle higher education altogether. Instead, we should engage the public in our research process, build a scientific practice representative of and accountable to the communities we serve, and interrogate the aims of our work by critically studying the history of science.

# 1. Introduction

After years spent enduring a pandemic, hiring freezes, and stalled projects, US scientists now face a rising tide of popular skepticism in academic institutions, and a White House-led effort to completely unravel over eighty years of close partnerships between the federal government and researchers in medicine, engineering, and the basic sciences. The dust plume around Washington, D.C. has yet to settle, the implications of the barrage of Oval Office policy actions are already clear: mass firings at national science laboratories and funding agencies,[1,2] interruptions and terminations of federal science grantmaking,[3,4] and the evisceration of training and fellowship programs for the next generation of students and scientists.[5,6]

Along with deep cuts to the American research enterprise, the emerging policy agenda has sought to limit the scope of scientific inquiry in the US on ideological grounds: projects that touch on disfavored subjects[7] such as race,[8] gender,[9] online misinformation,[10] and climate change[11] have been cancelled;[12] demands have been made for government control over university governance and hiring;[13,14] and, inquiries have been directed at academic journals concerning whether they are "partisan in various scientific debates."[15,16] These disruptions threaten research and scientific progress in the United States, and — taken together — constitute a wholesale campaign to undermine the independence of the American academic institutions and diminish their role in the public sphere.

Amid this crisis, many scientists and their institutions have faced a common question: how should we respond?

The instinctive response of many research institutions and universities has been a retreat from public visibility in hopes that maintaining an inoffensive posture and refocusing on "pure" scholarship will help them weather these political winds. Scrubbed diversity and outreach programs,[17] renewed commitments to "institutional neutrality,"[18] and tightened campus speech and protest codes[19] have been hallmarks of university administrators worried about angering donors, or, worse, attracting the unwanted attention[20] of Trump's political appointees and populist congressional allies in Washington.

While this strategy might make tactical sense, as a long-term vision it neglects the structural reasons for the current predicament. Under-engagement, not over-engagement, with the concerns of the American people has shattered their confidence in scientific institutions. Indeed, the language adopted by the most strident antagonists of American universities and the scientific establishment, including Trump administration officials[21] and policy whisperers,[22] paints academies as closed, monocultural, and distant from the American public. There is an element of truth in many of these critiques. But the policy solutions that have been proposed and enacted — axing STEM education[23] and outreach,[24] canceling projects that study and track disparities in health[25,26] and education,[27,28,29] and defunding initiatives to increase the diversity of the academy[30,31] — will only undermine the academy and exacerbate its shortcomings.

Reform is needed. But it requires deliberate effort from within the scientific community, not concessions to external pressure, especially when that pressure aims to stifle rather than encourage free inquiry. And a defense of the academy must tackle not just the present crisis, but also the underlying reasons for its emergence: rising antipathy toward scientific methods and conclusions, pent-up frustration with a perceived class of elite and unaccountable experts, and growing doubts about the value of academic institutions and education. These trends, if unabated, pose an enduring, existential threat to the ecosystem of scientific research in the US and the relationship with the American public that empowers and funds it.

## 2. The COVID-19 pandemic and the rejection of scientific authority

In spring 2020, as SARS-CoV2 spread across the globe, nations locked down to slow transmission, ration the limited number of hospital beds, and protect the vulnerable until a vaccine could be developed.[32] Two stories were told about the American scientific community and its response to the COVID-19 pandemic in the years that followed.

The first is an account of technical capability and scientific ingenuity: scientists at Moderna and the National Institutes for Health (NIH) designed a vaccine[33] for SARS-CoV2 within days of researchers in China first publishing the viral genome in 2020,[34,35] and high-efficacy vaccines[36,37,38] from both Pfizer and Moderna were administered in the US before the end of the year, almost nine months earlier than epidemiologists imagined.[32] Decades of basic science research,[39] much of it publicly funded via federal science agencies,[40] led to the mRNA vaccine technology that made the swift emergency response possible.

The second story is one in which scientists and health officials were not protagonists but instead unsympathetic architects of restrictive lockdowns and manipulators who used their authority and expertise to advance a political agenda. The mainstreaming of deep-seated skepticism towards the medical and scientific communities has overshadowed the extraordinary chronology of early successes in implementing public health measures, developing a vaccine, and averting possibly millions of American deaths.[41]

It is true that scientific leaders were not faultless in responding to the pandemic. In their efforts to anticipate and manage public sentiment, health officials sometimes spoke with too much certainty, giving people reason to doubt their claims when guidance shifted.[42] As late as March 9, 2020, Anthony Fauci, director of the National Institute of Allergy and Infectious Diseases, reassured Americans that the novel coronavirus was not a major threat in the United States.[43] This stance was quickly reversed: on March 13, the Trump administration declared a nationwide emergency and issued travel bans while states began implementing social isolation protocols.[44] Guidance on masking similarly reversed course within two weeks in March, with Fauci and the US Surgeon General first insisting that masks were unnecessary for people without symptoms before advising the use of face coverings in all public spaces.[45] Recent reporting also suggests that some prominent scientists closed ranks early in the pandemic to limit discussion of the possibility that SARS-CoV2 emerged from a lab and to direct research efforts away from this hypothesis. This reluctance left the public conversation concerning the virus's origins to unfold on the terrain of science skepticism and Sinophobia.[46] Contradictory directives given by the leaders of the American scientific establishment fostered skepticism and distrust of science in the public, which was suffering from not only the pandemic health crisis, but also from the additional disruptions caused by lockdowns, remote work, and remote learning.

Failures of communication did not create science skepticism, but they took place at a critical point for the erosion of trust in experts. Nonpharmaceutical interventions such as lockdowns initially had a broad mandate, with forty-three of America's fifty governors issuing stay-at-home orders.[47] But, as the pandemic wore on, rightwing politicians successfully cast lockdown policies as tools of government encroachment on personal freedom,[48] while politicians on the left continued to cite scientific data and expert opinions to justify restrictions that they themselves sometimes flouted.[49] As a result, scientific expertise became associated with political partisanship rather than objective fact. As misinformation spread, anti-science sentiment crystallized as a rejection of a class of epistemic elites — scientific experts and the mainstream media that putatively dictate the lives of ordinary citizens. These beliefs often coincided with conspiratorial ideas, like the notion that vaccines were used to implant microchips to track individuals.[50] When President Biden took office in the midst of these divisions, he fell short of campaign promises to ensure reliable access to COVID testing,[51,52] and the sense of the pandemic as an ongoing public health emergency diminished despite the fact that more Americans would die of COVID-19 after vaccines first became available than had died beforehand.[53]

The consequences of the split narratives of the pandemic were revealed by a Pew Research Center poll that showed the public's trust in science has fallen since 2020, particularly among right-leaning voters.[54] As of November 2024, over a third of right-leaning voters have "not too much/none at all" confidence in scientists to act in the best interests of the public — more than double the proportion in the first months of the pandemic — and a disturbing 36 percent of respondents across the political spectrum believe research scientists "don't pay attention to the moral values of society." In a stark illustration of how once-fringe positions on vaccines have gone mainstream, Robert F. Kennedy Jr., who has consistently spread doubt about the safety of the measles vaccine, is now the Secretary of Health and Human Services.[55] In 2025, for the first time in a decade, there have been outbreaks of measles in the United States,[56] a disease associated with as much as half of all childhood mortality before vaccines were developed.[57] Nevertheless, in June, Kennedy fired all seventeen members of the CDC's vaccine advisory board and began replacing them with anti-vaccine advocates.[58]

The majority of Americans still believe scientists work in the public interest, but the growing consumption of anti-science rhetoric peddled by influencers and politicians with a personal stake in distorting the truth is alarming. Meanwhile, the physical sciences remain siloed spaces, separated from the humanities, the social sciences, and the public. It is tempting to imagine that objective methods enable science to operate in the absence of trust, but if we critically reflect on our own thinking, we can see that is not the case. While we may understand how to design experiments or interpret statistical data, for fields outside our specialties we still depend on the scientific establishment and peer review process to produce knowledge that exceeds our ability to verify it. Data is not persuasive in and of itself, and knowledge that needs experts to produce it also demands that non-experts trust it. Science ought to be communicated as an ongoing process rather than a set of fixed truths, and we should not assume the public will be content to trust the outcome of any inquiry that denies room for their doubts.

### 3. The institutional embrace of DEI and the backlash

The pandemic saw another profound shift in public life: societal inequalities came to the fore of American consciousness. As COVID-19 spread in the United States, hate crimes against Asians surged, and, in response, anti-Asian racism was denounced online under the popular hashtag #StopAsianHate.[59] Then, in the summer of 2020, the murder of George Floyd sparked global protests against police brutality and excessive use of force against Black Americans.[60] The consequent groundswell of support for the Black Lives Matter movement prompted widespread reckonings with structural inequity in workplaces, schools, and municipal life. Under mounting public pressure, numerous corporations instituted new diversity, equity, and inclusion policies.[61] Discussions of these policies, grouped under the DEI acronym, as well as their underlying principles, pervaded not only the corporate world but also universities and government offices.

As public support fueled the expansion of DEI, concomitant criticism denounced new policies as unjust and anti-meritocratic. The outlines of this debate — as well as the tendency for its participants to draw opposite conclusions from the same facts — is illustrated by the story of the National Football League's (NFL's) "Rooney Rule," first adopted by the league in 2003.[62] The Rooney Rule requires that a minority candidate be included in the interview process for important jobs, and, in the post-2020 DEI surge, numerous other organizations adopted similar hiring policies, citing the NFL as inspiration. (Many of us in academia may recognize Rooney Rule-flavored approaches in hiring and admissions.) The Rooney Rule became popular because it was simple, but its simplicity also makes it easy to circumvent: interviewing a minoritized candidate does not have to influence hiring choices at all. At the same time, the policy's visibility allows opponents of DEI to argue that race and gender play too large a role in hiring decisions. As a result, the NFL is facing Rooney Rule-related lawsuits from both proponents and opponents of DEI. America First Legal, a rightwing nonprofit (founded by White House arch-strategist Stephen Miller), alleges that the Rooney Rule is racist and illegal because it privileges minoritized candidates.[63] Meanwhile, a lawsuit from Black head coach Brian Flores alleges that he was racially discriminated against, citing text messages indicating the New York Giants had decided to hire a white candidate days before they interviewed Flores[64] — in other words, that the team's compliance with the Rooney Rule was all pageantry.

Whether inclusion threatens meritocracy was a controversial subject on college campuses well before 2020, as exemplified by the debate over affirmative action. While affirmative action dramatically expanded access to American universities to underrepresented minorities,[65] like the Rooney Rule in the NFL, it was an imperfect solution to a deeply rooted problem. Even while cultivating more racial representation, elite universities continued to admit classes of students that were decidedly not diverse in terms of socioeconomic background, composed of as many students from the top one percent of family incomes as the bottom sixty percent.[66] This inequity was worse for minoritized students: admitted students from underrepresented racial groups were disproportionately from high income families, even more so than their non-minoritized peers,[67] meaning that schools were diversifying racially without necessarily increasing class mobility. This phenomenon can be described as a kind of "elite capture," in which policies intended to improve equity for a group of people tends to be taken over by elites within the group. In 2022, the philosopher Olúfẹ́mi Táíwò used the concept of elite capture to account for why DEI movements and identity politics often privilege the voices of educated or wealthy elites within racially minoritized groups.[68] Táíwò points out that minoritized people who make it into positions of power often assimilate to the values and standards of people already in power, so systemic change resulting from increased diversity can be meager. Racism disadvantages people of color even when they achieve high socioeconomic status, justifying the consideration of race rather than solely class,[69] but many DEI initiatives that were intended to check boxes and improve optics enacted a form of identity politics that only served elite interests.

While the effectiveness of affirmative action was curtailed by its uneven application across class divisions, criticisms of the practice suffered from the same bias, dominated by racial elites among the middle-class applicants who correctly sensed that the admissions process was stacked against them. In *Students for Fair Admissions v. Harvard*, the legal case that ultimately undid affirmative action, a group of Asian and white plaintiffs alleged that the admissions process privileged Black and Hispanic applicants.[70] For top American universities with notoriously opaque "holistic" admissions policies, wealthy and well-connected applicants with access to resources such as private college counseling have a profound advantage in the admissions process.[71] Even so, popular discourse has long fixated on allegations of racial discrimination in admissions instead of socioeconomic inequity. The importance of the *Students for Fair Admissions* case underscores this fact. If we take a step back, we can see that these heated debates over discrimination in admissions policies only matter for the select few "top" universities that admit fewer than twenty percent of their applicants. The majority of universities in the United States accept the majority of their applicants.[72] It is only if we take for granted that the most exclusive universities are necessarily the most important ones, responsible for reproducing the socioeconomically elite, that racial achievement in college admissions must seem zero-sum.

Debates over admissions policies at elite colleges should not overshadow the inequalities in access to education that lead various populations to arrive at the college application process with unequal qualifications. If we return to the earlier example of the NFL, we can see that the league did diversify over the past decade. Nearly half of its current coaching staff are people of color, and at the beginning of the 2024 season, fifteen of the thirty-two starting quarterbacks were Black.[73] The lesson we can learn from the NFL is that diversification ultimately happens when underrepresented people can perform at the highest levels — the number of Black quarterbacks did not increase because of hiring quotas or because teams wanted to remediate a history of racism, but because the Black quarterbacks in the league are very good.[62] Similarly, diversity will increase in academia when historically excluded people are supported throughout their careers and the barriers to their success are removed. This endeavor will require more than simply accounting for the race of candidates in hiring or admissions processes (though these policies are a starting point for improving representation). Pursued in this way, initiatives to increase diversity pose no threat to meritocracy, but in fact broaden the pool of talented candidates.

As skepticism of scientific experts grew during the COVID-19 pandemic, so did skepticism of the elite institutions that housed and produced them. Today, criticisms of DEI are being wielded as evidence of general institutional mismanagement, facilitating attacks on higher education as a whole. The 2023 Supreme Court ruling on *Students for Fair Admissions v. Harvard* now forms the legal basis for the dismantling of DEI programs in schools.[74] And the protests against the war on Gaza that roiled universities in the spring of 2024 produced a convenient ground for the right wing to curtail free speech on campuses and revoke student visas in the name of national security.[75] Campus antisemitism is the explicitly given reason for which the government has withheld scientific funding from Harvard, Columbia, and other elite institutions, and Civil Rights law is being wielded as a cudgel to back up these threats.[76]

**4. Defending science against bad faith attacks**

The current attacks on scientific funding came about after a racially and economically elite coalition with a sea of grievances against the existing establishment took power alongside Donald Trump in January 2025. Within this coalition, some wanted to weaken government institutions,[77] some took aim at universities and "woke" campus culture,[78,79] and others were antagonistic toward medical and scientific expertise and its claims to authority.[80] Universities are the nexus where these nominally anti-elitist goals intersect. Although attacks on academic institutions may have been hinted at in documents like Project 2025 (a policy wish list circulating around the 2024 Trump campaign that included a proposal to eliminate the Department of Education),[81] the swift and comprehensive dismantling of the American scientific infrastructure came as a surprise to almost everyone at the start of this year. Perhaps no individual faction of Trump's coalition would have stated an intention to halt scientific research entirely, but together they have settled on dismantling science as the best way to fulfill their disparate and mutual ambitions. Criticisms of science and DEI are based in some real weaknesses in their practice, but current attacks on universities are not grassroots movements to reform science — they are efforts by a different handful of elites to destroy higher education altogether, disguised as legitimate concern about elitism.

We should work to correct flaws in the practice of science and inclusion, but we cannot cave to the demands of the government in hopes of weathering the current hostility toward universities. Compliance will only further erode public trust in higher education by demonstrating that universities are indeed sycophants driven by the interests of the wealthy and powerful — and that policies focused on inclusion were only palatable when they were popular. While universities with large endowments have been able to mount legal opposition to the Trump administration and appeal to wealthy donors for additional support,[82] this model of resistance fails to map a complete path forward. A future in which knowledge generation is the province of a handful of wealthy universities and their donors can only be worse for intellectual freedom than the imperfect public funding system we had prior to this year. Rather than retreating from scrutiny, academics must win back public support. Doing so does not entail abandoning prior commitments to diversity or pandering to the interests of any elite coalition. Instead, it means continuing the work of building a diverse scientific community and accountable scientific practice so we can prove that our research expertise has not made us insulated elites. Public trust and accountability may seem like distant concerns for the physical sciences, but we hope to show that there is a necessary connection between diversity and accountability for basic science in particular.

## 5. Engaging the public in basic science

A tension resides at the heart of scientific practice. On the one hand, science must be autonomous, with research directions and methods dictated by experts and free from partisan influences. On the other hand, science must also demonstrate that it is accountable to the public interest. The federal government remains the largest source of basic research funding in the United States,[83] making scientific choices necessarily political choices about how to allocate taxpayer money.[84] In a truly democratic society, science that has lost its responsiveness to the public ought to lose the trust of the public and in turn its funding. The problem with the current attacks on higher education is not that they take dollars away from science but that they do so in service of an elite minority trying to police dissent. Although the actions of the Trump administration reveal an unmistakable connection between systems of power and the production of knowledge, that connection already existed and has simply become more apparent.

Scientific work necessarily conforms to the priorities of those who fund it.[85] In 2008, the NIH spent $5.7 billion funding research on cancer, Alzheimer's disease, and cardiovascular disease, which collectively affected 22 million Americans,[86] whereas that year only $347 million was spent globally toward research on neglected tropical diseases that affect over 1 billion people annually.[87,88] The people affected by neglected tropical diseases primarily live in developing countries where research funding is scarce, while those affected by geriatric diseases such as cancer and Alzheimer's primarily live in wealthy Western countries that control the overwhelming majority of global research funding (one World Health Organization report suggests that the NIH funded 88 percent of global biomedical research in 2022).[89,90]

Just as the countries with the power to disburse research funding have an exaggerated influence on which research questions get answered, so too does representation affect scientific inquiry within one country. A lack of representation in scientific practice has harmed neglected communities in the US: in the now infamous Tuskegee syphilis trials of the mid-twentieth century, Black study participants with syphilis were denied treatment for decades after one became available so scientists could observe the effects of the untreated disease.[91] And, as late as the mid-1990s, researchers at Johns Hopkins University knowingly exposed study participants (the vast majority of whom were Black single mothers and their children) to lead paint to study its relationship to childhood neurological disorders, despite many negative health effects of lead exposure already being established.[92] Many similar examples of unethical scientific research exist, and the victims are usually low-income people of color. Early clinical trials for hormonal birth control were conducted in Puerto Rico, where Hispanic women who participated in the studies were given extremely high doses of estrogen and progesterone to ensure its efficacy.[93,94] As a result, three of those women died in the process of developing a drug initially priced too high for them to have been able to afford. Communities historically underserved by the medical apparatus distrust the system with legitimate reason: they have disproportionately borne the risks and harms of scientific research. This history of neglect of women and ethnic minorities is not the animating force inside the current anti-science coalition, but the hostile feelings that motivate today's anti-science movement can be traced through similar narratives, such as the opioid crisis that was driven by deceptive marketing from Purdue Pharma and led to tens of thousands of deaths from addiction and overdose, chiefly among low-income white Americans in rural communities.[95,96]

Within this history, we can also find illustrations of how to regain public trust. When COVID vaccine distribution began in 2020, Black Americans were disproportionately hesitant to be inoculated.[97] But the disparity in vaccination rates for Black Americans closed after trusted community leaders conducted outreach in collaboration with doctors to address vaccine hesitancy by answering questions and assuaging concerns.[98,99] In the last forty years, community-based public health interventions have been shown to succeed in building trust with various communities, leading to positive health outcomes.[100] Until all communities can achieve access to the scientific establishment and thereby advocate their own interests, the distribution of benefits and harms from the impacts of research will continue to be inequitable. These inequities breed distrust, and in turn weaken support for science funding.

But conversations about community involvement in research have yet to take hold in the physical sciences, often dismissed with the reasoning that our work is too far removed from applications for such discussions to be relevant. In fields such as public health and environmental sciences, methods of community-based participatory research and co-production of knowledge have gained momentum as methods to involve stakeholders (populations that are most affected by the research) in the process of research.[101] In these methods, scientists communicate iteratively with stakeholder communities and include their feedback in decision-making. For basic science, a few degrees separated from direct applications, the identity of the stakeholder communities is less clear. Without knowing the future impacts of the research, we cannot always identify the people, if any, who will be affected by our work. Envisioning the possible impacts of a basic research direction may also require expertise and scientific understanding, making it difficult to involve non-experts in decision-making.

Basic science has the potential, however, to translate to world-altering technology in short spans of time. The Manhattan project provides a quintessential illustration of this scenario — only six years elapsed between the discovery of atomic fission and the development of the first atomic bomb.[102] Scientists quickly realized the destructive possibilities of fission, but still failed to protect vulnerable communities from the consequences of their research: aside from the development and deployment of the most lethal weapon in human history (the ethics of which are still contested), nuclear weapons research exposed numerous indigenous American communities to harmful radiation and radioactive pollution, and many of these communities suffer from elevated rates of cancer to this day.[103] CRISPR gene editing provides a contemporary example of a basic science finding with massive translational implications, and some scientists and public health experts have urged the use of community-based participatory research to ensure the consideration of vulnerable populations.[104] If technological advancement continues to accelerate, one can only expect such cases of rapid translation from basic research to applications to occur more frequently, making it increasingly important for physical scientists to engage in conversations about the long-term consequences of their work.

Furthermore, while the fact that basic research could affect anyone makes it harder to determine how to involve affected communities, it also makes it all the more important that we do. In the physical sciences, we can achieve the co-production of knowledge between experts and stakeholders by creating a research community representative of the general population. If everyone has equitable access to scientific training and can pursue a career in basic science that could culminate in conducting and directing their own scientific research, then the various interests of our population will naturally be represented. And if the composition of experts and stakeholders reflect one another — if the experts themselves are a representative set of stakeholders — then scientists can feel justified in arguing that expertise does not constitute membership in an elite group insulated from the concerns of the public. By similar reasoning, foreign students in the US comprise an essential means by which our country's research apparatus can incorporate the interests of the world at large, highlighting that we should resist the current administration's manipulation of foreign visas as leverage against universities.

Science should be a mechanism through which all communities can enact their interests and probe the questions that matter to them, and the practice of science should be governed by scientists and the diverse constituencies they represent. A small group of wealthy donors or powerful politicians should not be able to dictate acceptable research directions. No vision of science as a practice by which anyone can seek the truth is compatible with the withholding of grant money for research projects studying questions of gender, differential health outcomes in marginalized communities, or subjects with an even vaguely international focus.

We cannot ask communities to trust the scientific establishment without giving them access to its decision-making processes. Trustworthy science requires a diverse scientific workforce laboring to produce knowledge with equitable effects in society. The shortcomings of individual DEI policies do not demonstrate that we should abandon the work of inclusion altogether. Almost everyone practicing science today has benefitted from the long history of movements that struggle against hierarchies that limit access to educational and scientific achievement, be they the Civil Rights Movement or even class struggles against feudal systems — just over two hundred years ago, scientific research was primarily conducted by nobility.[105] We must continue this work, not because it is trendy, but because it benefits humankind, and because, we, as scientists are especially tasked with making our work and our methods accessible if we want to assume the mantle of objectivity.

## 6. Interrogating our aims

Enabling people from all parts of society to represent their interests by becoming scientists is not alone sufficient; we must also ensure that scientific education and practice does not distort the interests of its practitioners. As arguments about elite capture suggest, having a diverse workforce is not enough unless we also foreground the interests of those excluded from privileged decision-making spaces. While the scientific method has proven effective at generating knowledge, it is agnostic to how that knowledge is used. Although physics and chemistry can explain how the world works, they cannot explain what we should do with that knowledge. If the scientific establishment is to justify its continued status as a public good, it needs to look to history to better understand how science functions in relation to the public. One starting point is to include the history of science in our scientific curriculum. There is currently little discussion of the history of science and its social impacts in the physical sciences curriculum at the undergraduate and graduate levels. These settings will produce tomorrow's scientists. Excluding such topics signals that they do not constitute essential knowledge for a scientist, but if that illusion could once be comfortably maintained, the current crisis has dispelled it.

At UC Berkeley, the graduate chemistry curriculum was modified in 2022 to include a discussion-based course titled "Scientific Responsibility and Citizenship" that examined the inequitable distribution of the risks and benefits of chemistry research between different communities.[106] That work, undertaken by an author of this editorial, was evidence-based: the development of the course built on data from a departmental survey,[107] and the efficacy of the course was studied using chemical education research methods. The course suggested that conducting good research requires thinking about how it ultimately affects people. Students in the course learned about the contraceptive trials in Puerto Rico and the effects of the Manhattan project on indigenous Americans, as well as the geopolitics of rare earth mining and the regulation of environmental pollutants. Students discussed the challenges in ethical decision-making and the capabilities and limitations of scientists in these situations. Literacy in the history of science enables us to make more informed decisions in response to future predicaments. By prioritizing scientific citizenship in our curriculum, future scientists will become more adept at navigating the relationship between basic science and the public.

Students and faculty also appreciated opportunities to have moderated discussions of contentious topics. Discussion and disagreement comprise core components of research , as they are necessary for interpreting new results, and the course created a space for students to practice these skills. Critics of higher education often attack universities for a lack of "viewpoint diversity" — differences in political leanings and opinions among students and faculty.[108] Science cannot progress through consensus thinking, but we should not try to achieve diversity of thought by sorting college seniors or faculty hires by political party or forcibly determining which viewpoints should be present in our communities: real differences in opinion emerge spontaneously when researchers discuss contentious topics that lack well-resolved consensus stances. Critical reflection on the history and practice of science provides the means to cultivate viewpoint diversity. Likewise, empathetic disputes made possible by the curriculum itself can address concerns[109,110] that the ideological dominance of political correctness created an environment where people were afraid to say the wrong thing or that well-intentioned people who caused offense would face backlash disproportionate to the infraction. An atmosphere in which dialogue is restricted ultimately stifles progress, and a culture unforgiving of mistakes and disagreement is inhospitable to scientific thinking. We should instead foster spaces in which respectful disagreement, a necessary intellectual skill, can be learned and practiced.

Academic discourse in the physical sciences is dominated by discussions about how to improve the precision, versatility, and efficiency of our research methods. We urge the community to pay more attention to how scientific knowledge propagates through society and how it can be made to benefit the public. We believe doing so entails continuing the work of making our research communities representative of the public while also making science better able to reason about its practice and its purpose. You, our readers, are sure to have your own ideas. We encourage you to discuss those ideas in your community and with us. To oppose the coalition surrounding the Trump administration, we must form a coalition of our own to defend our work and our principles. Conversations about the social consequences of our work can bridge divides between academic subfields and connect with the broader public. Our coalition need not agree on every detail, but if we all believe in defending science and higher education, then we should start a critical dialogue about the role of science in society.

We propose these questions to begin that conversation:

- What do you think is the purpose of your work as a scientist?
    - How is it shaped by your values and worldview? What benefits do you hope will result from your research? What are possible risks or harms of the work? Who will be affected by your work?
    - How might you communicate your work to someone outside your field? To a non-scientist?
    - If you were to ask the same questions of your colleagues, how might their answers differ? Whose perspectives and interests may not be represented by you or your colleagues, but are still worth considering?

- How can we mitigate the harmful effects of our work?
    - The nature of basic research is such that the new technologies we develop can contain unknown risks. If we were to envision all the possible negative consequences of our work, how would we then practically integrate that knowledge into our research process? Some risks may be unavoidable, and if so, how can the burden of risk be fairly distributed?

- How do power structures influence our practice of science?
    - Power is configured along political, cultural, social, and economic axes. How do these structures determine what research we do, who gets to do it, and how the results are used? What can researchers practically do to uncover and counterbalance these structures when they are inequitable?
    - Historically, how has research in your field been funded and who has benefited from the technologies derived from developments in basic science?

We, the scientific experts, are the only ones capable of answering these questions for our work. We are capable of answering them at any career stage and any career path. And, if we wish to protect the practice of scientific research in the years to come, we must answer them.

**Author Information**

K.T.X. wrote and researched the initial draft. T.L.A. and P.Y. reviewed and edited the argument. All authors participated in revising and finalizing the manuscript.

Corresponding author: K.T.X. (ktx@berkeley.edu)

**Acknowledgements**

The authors thank numerous readers of drafts of this article prior to submission. These readers include Camila A. Suarez, Robert G. Bergman, M.I., S.L., and others who have asked not to be named.